\documentclass[fleqn,12pt,twoside]{article}
\usepackage{espcrc1}
\usepackage{epsfig}

\usepackage{graphicx}
\usepackage[figuresright]{rotating}

\newcommand{\AmS}{{\protect\the\textfont2
  A\kern-.1667em\lower.5ex\hbox{M}\kern-.125emS}}

\title{Taming the BFKL Intercept via Gluon Saturation}

\author{Alexander Kovner\address[MCSD]{Department of Mathematics and Statistics,University of Plymouth,\\
2 Kirkby Place, Plymouth, PL4 8AA, U.K.}
and Urs Achim Wiedemann\addressmark[URS]{\thanks{presenting author}
\address[URS]{
Theory Division, CERN, CH-1211 Geneva 23, Switzerland}}
}       
\begin{document}

% typeset front matter
\maketitle

\begin{abstract}
We show that the inclusion of parton density effects in 
the perturbative small-x evolution reduces the strength 
of the powerlike growth of total hadronic cross sections. 
\end{abstract}

\section{Introduction}

At center of mass energies above $\sqrt{s} \approx 100$ GeV,
total hadronic cross sections $\sigma$ show a continuous rise which persists
up to the highest energies explored so far and which is consistent 
with a gentle powerlike growth, 
\begin{equation}
  \sigma \propto s^{\alpha_P}\, ,
  \qquad \hbox{\rm where} \qquad \alpha_P=0.08\, .
  \label{eq1}
\end{equation}
However, at asymptotically high $\sqrt{s}$, hadronic cross sections 
satisfy the unitarity or Froissart bound, 
$\sigma \leq \frac{\pi}{m^2} \ln^2 \frac{s}{m^2}$
where $m$ is the smallest mass in the theory. Thus, the power law 
growth (\ref{eq1}), parametrized by the soft pomeron intercept $\alpha_P$,
can be valid only in a preasymptotic, though possibly large, kinematical
regime. So far, this growth is not understood in the context of QCD.
Indeed, perturbation theory in the LO BFKL framework does result in
a powerlike growth 
\begin{equation}
  \sigma \propto s^{\omega}\, ,
  \qquad \hbox{\rm where} \qquad \omega = \frac{\alpha_s}{\pi}
  N_c 4 \ln 2 \gg \alpha_P\, ,
  \label{eq2}
\end{equation}
but the BFKL intercept $\omega$ is much larger than the experimentally
observed one. Inspecting at large rapidity 
$t = \ln \frac{s}{m^2}$ the analytically known dependence of the
BFKL density $\Phi(b,k,t,k_0)$ of gluons of momentum $k$ at impact 
parameter $b$ originating from a gluon of momentum $k_0$ at initial
rapidity $t_0$, two different perturbative (non-unitary) growth 
mechanisms can be identified in the BFKL calculation:
\begin{enumerate}
  \item \underline{Growth in density:} For $b < b_{\rm diff} \simeq
    e^{\sqrt{\alpha_s}t}$, one finds $\Phi(b) \sim e^{\omega t}$.
    For BFKL, this is the {\it dominant growth mechanism}, 
    resulting in (\ref{eq2}).
  \item \underline{Growth in impact parameter space:} For 
    $b_{\rm spread} = e^{\epsilon \alpha_s t}$, one finds
    $\Phi(b_{\rm spread}) \sim 
    e^{\left(\omega - \epsilon^2 \alpha_s\right) t}$. For
    BFKL, this is a {\it subleading growth mechanism}.
\end{enumerate}
Here, we review arguments that including ``gluon saturation'',
i.e. parton density effects, in the LO perturbative QCD calculation
allows to regulate the dominant growth in density, thus taming
in comparison to BFKL the non-unitary powerlike growth to $\sigma \propto
s^{\epsilon \omega}$. This provides a perspective for a perturbative 
calculation of the soft pomeron intercept.

%%%%%%%%%%%%%%%%%%%%%%%%%%%%%%%%%%%%%%%%%%%%%%%%%%%%%%%%%%%%%%%%%%%%%%
\section{The dipole-hadron total cross section}
For an explicit discussion of gluon saturation effects, we take 
recourse to a toy model of a hadronic collision. At initial 
rapidity $t_0$, we consider a perturbative $q\bar{q}$-dipole
with quarks at transverse positions $x$ and $y$
colliding with a hadronic target. The dipole scattering probability
is $N(x,y)$ [we also write $N(b=\frac{x+y}{2},r=\frac{x-y}{2})$],
and the total cross section is obtained by integrating over the
impact parameter, $\sigma(t_0) = 2 \int d^2b\, N(b,r)$. To calculate
the rapidity dependence of the cross section, one may work in the
target rest frame where $t > t_0$ corresponds to boosting the
projectile. This boost generates additional gluons which in the
large $N_c$-limit translate into additional dipoles at different
transverse positions. Pictorially:
\begin{eqnarray}
\epsfxsize=14.7cm 
\centerline{\epsfbox{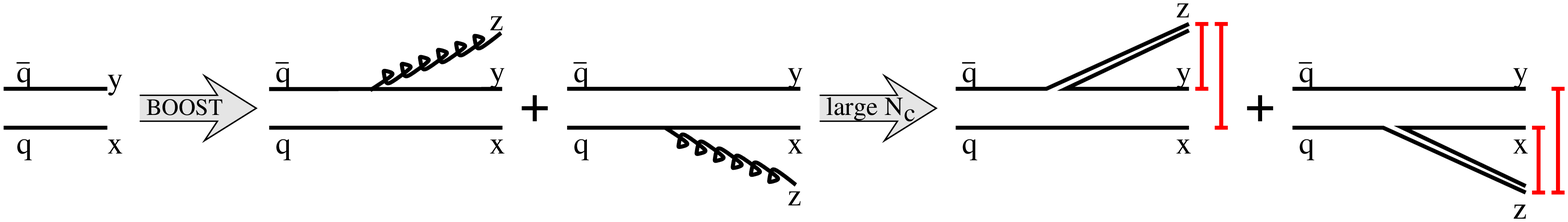}}
\vspace{-1cm}
\nonumber
\end{eqnarray}
Calculating the interaction of this boosted
dipole wave function with the hadronic target and taking the 
derivative with respect to the phase space opened up by the boost,
one regains the Balitsky-Kovchegov equation~\cite{bk}:
\begin{eqnarray}
  {d\over dt}N(x,y) = \bar{\alpha}_s\,
    \int \frac{d^2z}{2\pi}\, {(x-y)^2\over(x-z)^2(y-z)^2}
  [N(x,z)+N(y,z)-N(x,y)-N(x,z)N(z,y)] \, .\label{eq3}
\end{eqnarray}
This non-linear evolution equation differs from the linear BFKL
equation by the addition of the last term $N(x,z)N(z,y)$ only.
The physics contained in this modification of the BFKL equation
is that of a {\it double counting correction}. Namely, for a
projectile consisting of two dipoles $(x,z)$ and $(y,z)$, the
scattering probability is the sum of the individual scattering
probabilities minus the probability that both dipoles scatter:
\begin{equation}
  N(\lbrace x,z\rbrace ;\lbrace y,z\rbrace ) = N(x,z)+N(y,z)-
  N(x,z)\, N(z,y)\, . \label{eq4}
\end{equation}
The remaining fourth term $N(x,y)$ in (\ref{eq3}) is a virtual
correction required by the proper normalization of the projectile
wavefunction $\vert q\bar{q}\rangle + \vert q\bar{q}g\rangle$. 
In conclusion, the BK evolution equation (\ref{eq3}) can
be viewed in the target rest frame as BFKL evolution of the 
boosted projectile but supplemented
by the double counting correction for the interaction of the
evolved projectile wavefunction with the target. The swelling of the
BK dipole projectile wavefunction in impact parameter space is thus
known from BFKL. The density of dipoles of size $r$ 
at $b$ obtained by boosting the initial dipole of size $r_0$ to
rapidity $t$ takes the form
\begin{eqnarray}
  n(r,b,r_0,t)={32\over r^2} \frac{\ln {16b^2\over r_0 r}}{(\pi
  a^2t)^{3/2}} \exp\left[ \omega t-\ln {16r^2\over r_0 r} -{\ln^2
  {16b^2\over r_0 r}\over a^2t}\right]\, .
  \label{eq5}
\end{eqnarray}
To determine the growth of the total cross section from this density,
we consider the smallest dipole for which the scattering probability
$N(r,b)$ is of order one for $b<R_0$:
\begin{eqnarray}
\epsfxsize=14.7cm 
\centerline{\epsfbox{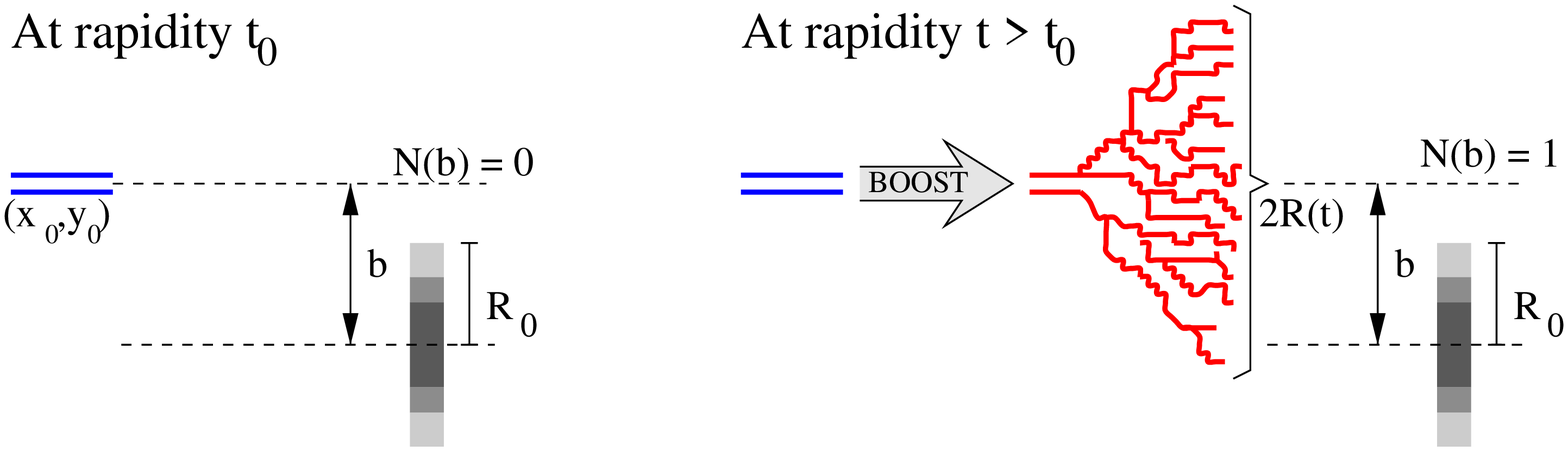}}
\vspace{-1cm}
\nonumber
\end{eqnarray}
If after boosting the density of dipoles at $b$ becomes unity, 
then the scattering probability is of order unity as well. This
establishes the non-unitary growth~\cite{kw01,kw02} 
\begin{equation}
  \sigma(t) = \pi R^2(t,Q^2=1/r^2)\quad \hbox{with}\quad 
  R^2(t,Q^2)={r_0\, r\over 16} \exp\left[ \epsilon t\right]\, .
  \label{eq6}
\end{equation}
From (\ref{eq5}),
one finds for the BK intercept $\epsilon = \frac{\alpha_s N_c}{\pi}
7\, \zeta(3) \Big [-1+\sqrt{1+ {8\ln 2 / 7 \zeta(3)}} \Big ]$. This
expression is parametrically correct but numerically questionable
since (\ref{eq5}) is based on a saddle point approximation valid
for sufficiently small impact parameter only. Still, the numerical value 
$\epsilon = 0.87\, \omega$ is compatible with a numerical result of 
G. Salam~\cite{salam} who finds a non-unitary growth of total cross 
sections with $\epsilon \approx 0.75\, \omega$ in a simulation which 
encodes essentially the same physics as (\ref{eq3}).

%%%%%%%%%%%%%%%%%%%%%%%%%%%%%%%%%%%%%%%%%%%%%%%%%%%%%%%%%%%%%%%%%%%%%%
\section{Saturation without Unitarization: alternative derivations}

The above derivation was given in the target rest frame where the
evolution resides in the projectile wavefunction. The same result
can be obtained~\cite{kw01,kw02} within Weigert's formulation where the
evolution is ascribed to the target. This latter argument can be 
viewed as establishing the non-unitary growth of the hadronic cross  
section for a coloured scattering probe. Moreover, it was 
shown~\cite{Kovner:2002yt} that the total cross section violates
unitarity for a colourless projectile as soon as it does so for
a coloured one. These three arguments are independent of each other
but consistent with each other and with Ref.~\cite{salam}. They 
establish that while the BK equation ensures {\it saturation} of 
the scattering probability locally in impact parameter space 
(i.e. $N(b,r) < 1$ for all $b$), 
it violates {\it unitarization} of the total cross section~\cite{kw01}.

%%%%%%%%%%%%%%%%%%%%%%%%%%%%%%%%%%%%%%%%%%%%%%%%%%%%%%%%%%%%%%%%%%%%%%
\section{Beyond the BK equation (BBK)}

In the target rest frame, the only density effect included in the
BK equation (\ref{eq3}) is the nonlinear double counting correction 
in (\ref{eq4}). This correction sets in as soon as the {\it number}
of gluons in the projectile state becomes large, even if the gluonic
{\it density} may still be small. In addition, as soon as the partonic
density in the projectile reaches a critical value of order $1/\alpha_s$,
one expects wavefunction saturation effects which further tame the 
growth of density in the projectile and are not included in (\ref{eq3}).
The rapidity scales for these different effects are:\cite{kw01}
\begin{equation}
  t_{\rm BFKL} \propto \frac{1}{\alpha_s\, N_c}\, ,\qquad
  t_{\rm BK} \propto \frac{1}{\alpha_s\, N_c\, \ln(R_0/r_0)}\, ,\qquad
  t_{\rm BBK} \propto \frac{1}{\alpha_s\, N_c\, \ln(1/\alpha_s)}\, .
\end{equation}
At rapidity $t_{\rm BFKL}$, the BFKL equation violates unitarity
due to the growth in density. From (\ref{eq6}) one finds that
at rapidity $t_{\rm BK}$, the BK equation violates unitarity for 
a (small) dipole of initial size $r_0$ incident on a sufficiently
extended target of size $R_0$. This violation is due to the growth 
of the projectile in transverse size. The applicability of the BK 
evolution, however, crucially depends on the nature of the target. 
If the target is thick enough, so that the scattering probability 
is parametrically larger than $\alpha_s$ [for a large nucleus of 
atomic number $A$, it is $O(A^{1/3}\alpha_s)$], and if the
target is wide enough, so that saturation occurs before the
projectile radius swells beyond that of the target, then there is
an intermediate regime $t < t_{\rm BK}$ in which BK applies. However, 
if the target is a nucleon, neither one of these conditions is satisfied. 

Wavefunction saturation effects should set in when the interaction 
probability of a dipole of size $r$ with another
dipole of similar size is of order unity, $\alpha_s\, r^2\, n(r) \sim 1$.
Taking the density growth exponential in rapidity, 
$r^2\, n(r) = e^{\omega t}$, this indicates that saturation effects
beyond those included in the BK equation become relevant for 
$t > t_{\rm BBK}$.  

It is reasonable to expect that projectile wavefunction saturation 
effects further diminish the intercept parameter of total cross sections
beyond the exponent of the BK equation. The reason is that a tamed 
growth of the density in the projectile due to dipole interactions
will result in a tamed growth in impact parameter. This suggests 
the hierarchy
\begin{equation}
  \omega_{\rm BFKL} > \omega_{\rm BK}
  > \omega_{\rm BBK}\, .
  \label{eq7}
\end{equation}
Thus, one is led naturally to conjecture that 
a non-linear generalization of the BFKL equation which goes beyond
BK by taking into account wave function saturation effects 
(``Pomeron loops'') provides a perturbative framework for the 
calculation of the soft Pomeron intercept, $\omega_{\rm BBK} = \alpha_P$.
However, this requires that the perturbative growth mechanism occurs
in the hadron on a scale smaller than the confinement scale at which
the Froissart bound can be expected to arise as a consequence of
non-perturbative physics. As explained in Ref.\cite{kw02},
a picture of the proton as consisting of three small black discs 
associated with the valence quarks is consistent with what is known
about hadronic cross sections and allows for this possibility.

%%%%%%%%%%%%%%%%%%%%%%%%%%%%%%%%%%%%%%%%%%%%%%%%%%%%%%%%%%%%%%%%%%%%%%%%%%%%

\end{document}